\documentclass[10pt,a4paper]{article}

\usepackage[none]{hyphenat}
\usepackage[title]{appendix}
\usepackage{graphicx}
\usepackage{adjustbox}
\usepackage{multirow}
\usepackage{float}
\usepackage{url}
\usepackage{amsthm}
\usepackage{amsmath}
\usepackage{amssymb}
\usepackage{colortbl}
\usepackage[utf8]{inputenc}
\usepackage[T1]{fontenc}
\DeclareUnicodeCharacter{00A7}{\S}
\DeclareUnicodeCharacter{0394}{\ensuremath{\Delta}}
\usepackage{latexsym}
\usepackage{bm}
\usepackage{booktabs}
\usepackage{array}
\usepackage{lmodern}
\usepackage{placeins}
\usepackage{hyperref}
\hypersetup{hidelinks}
\usepackage{listings}
\usepackage{color}

\definecolor{dkgreen}{rgb}{0,0.6,0}
\definecolor{gray}{rgb}{0.5,0.5,0.5}
\definecolor{mauve}{rgb}{0.58,0,0.82}

\graphicspath{{./}{./Figs/}}
\DeclareGraphicsExtensions{.pdf,.png,.jpg}

\title{An Elo-based Rating System for Topcoder\\Single Round Matches}

\author{
Fred Batty\\
\texttt{fbatty@gmail.com}
}

\date{2026-02-21}

\begin{document}
\sloppy
\maketitle

\begin{abstract}
This paper presents an Elo-based rating system for programming contests, specifically Topcoder's Single Round Matches (SRMs).
We introduce a logarithmic rank-based performance metric that allows single-round, multi-player contest results to be incorporated into an Elo-style continuous rating framework.
Model parameters and adjustment factors are calibrated empirically by minimizing absolute prediction error over historical data, accounting for experience level, initial ratings, and competition characteristics.
The resulting system demonstrates improved rank predictions and rating progressions consistent with natural skill development over player careers.
\end{abstract}


\section{Introduction}

Single Round Matches (SRMs) were regular programming contests organized by Topcoder~\cite{TC} from 2001 to 2024.
Topcoder developed its own rating system, a model of ranking and volatility~\cite{TCRating}.
Players sometimes asked: What would SRM ratings be if they were Elo-based?
This research addresses that question by developing and evaluating an Elo-based rating system for Topcoder SRMs.

The Elo rating system is a measurement of performance, developed in 1959 by Arpad Elo~\cite{Elo}.

\begin{itemize}
  \item The Elo system is based on a Thurstonian model, where individual performances fluctuate from ratings following a statistical process, characterized by the win rate equation:
  \begin{equation}
  E = \frac{1}{1 + 10^{-D/400}}
  \end{equation}
  where $E$ is the expected score and $D$ is the rating difference between two players.
  
  \item A defining requirement of the Elo system is the integrity of the rating scale: ratings of the same numerical value should represent the same level of proficiency over time~\cite{Elo, FIDE}.
  
  \item Proficiency denotes an absolute level of performance, skill or ability, independent of other competitors' performances.
\end{itemize}

There is no standard method for applying Elo ratings to single-round, multi-player contests, where the observed data consist of a ranking rather than a set of pairwise results.
Since Elo ratings operate on scores---scalar quantities measuring performance---the ranking must be translated into a score-equivalent measure before it can enter the rating formula.
We derive such a quantity by counting the number of one-on-one wins implied by a given rank in an equivalent tournament.
Consequently, an n-player game can be rated like a two-player game.

We adapt the rating change formula for Topcoder SRMs with the following objectives:
\begin{itemize}
  \item Improve prediction of players' future performances
  \item Estimate player proficiency over time
  \item Adhere to Elo rating system principles
\end{itemize}

\section{Performance}

\subsection{Rank performance}

To adapt the Elo system for multi-player SRMs, we must assign a score to each rank. We therefore ask:

\begin{quote}
If the winner of a chess game scores one point, how many points should a player ranked $r$ in a round of $n$ players score?
\end{quote}

We answer this by counting wins in an elimination tournament.
Let $\mathit{RP}(n, r)$ denote the score assigned to rank $r$ among $n$ players.

In a tournament of $2^k$ players, the winner must win $k$ one-on-one rounds, so we have:
\[
\mathit{RP}(2^k, 1) = k
\]

The runner-up, having lost only the final round, has one fewer win than the winner:
\[
\mathit{RP}(n, 2) = \mathit{RP}(n, 1) - 1
\]

In other situations, we observe:
\[
\begin{gathered}
\mathit{RP}(n, n) = 0 \\
\mathit{RP}(2 n, n) = 1 \\
\mathit{RP}(n, r) = \mathit{RP}(n, 1) - \mathit{RP}(r, 1)
\end{gathered}
\]

These conditions uniquely determine the rank performance equation:
\begin{equation}
\mathit{RP}(n, r) = \log_2{n} - \log_2{r}
\end{equation}

This converts rankings into score-equivalent performance values suitable for Elo rating updates.

\subsection{Expectations}

We implement the standard Elo win rate equation~\cite[§8.73]{Elo}\cite{FIDE}.
A rating $R_j$ outperforms a rating $R_i$ with probability $w_j$ given by:
\begin{equation}
w_j = \frac{1}{1 + 10^{(R_i - R_j)/400}}
\end{equation}

This is also known as the Bradley-Terry model.

Following SRM convention, new players start at rating 1200.

\subsection{Ties}

In programming contests like SRM, ties typically indicate a limitation of the problem set or scoring rather than truly equal performances of the tied players.
We want ties to not adversely affect the ratings.

We experimented with several accounting methods and found the most effective is to split ties equally in both actual and expected ranks, assigning 0.5 to each tied position.
Thus, we split ties equally.

\subsection{Round performance}

We have the results of a round, which may include multiple divisions and ties.
We consider the results of each division separately.
The result of a round is a list of raw scores $S$, where $s_i$ is the score achieved by player $i$.

We compute rank and expected rank for each player $i$ as follows:
\begin{equation}
\begin{aligned}
r_\mathit{ties} &= \frac{1}{2}|\{s_j \in S: j \neq i, s_j = s_i\}|, \\
r &= 1 + |\{s \in S: s > s_i\}| + r_\mathit{ties}, \\
\hat{r} &= 1 + \sum_{j: s_j \neq s_i}w_j + r_\mathit{ties}.
\end{aligned}
\end{equation}

The net performance of a player in the round is the difference between actual and expected rank performance:
\[
P = \mathit{RP}(n, r) - \mathit{RP}(n, \hat{r})
\]

Simplifying, we obtain:
\begin{equation}
P(\hat{r}, r) = \log_2{\hat{r}} - \log_2{r}
\end{equation}

The performance $P$ equals the number of wins above or below expectations in an equivalent tournament.

Thus, the n-player game can be effectively rated as a two-player game using the following formula:
\[
\Delta R = K \cdot P
\]
where $K$ is the rate of rating change, to be determined.

\subsection{Accuracy}

To evaluate the rating system, we measure how accurately it predicts contest outcomes.
Given expected and observed ranks $(\hat{r}, r)$, we define the prediction error as the absolute difference in rank performance:
\begin{equation}
E(\hat{r},r) = \left| \log_2{\hat{r}} - \log_2{r} \right|
\end{equation}

Our primary accuracy metric is the average error for all participants in all rated rounds, denoted $L_1$.
This metric measures how accurately the expected ranks match the observed performances:
\begin{equation}
L_1 = \frac{1}{n} \sum_{i=1}^{n} \left| \log_2{\hat{r_i}} - \log_2{r_i} \right|
\end{equation}
where $r_i$ and $\hat{r}_i$ are the actual and expected ranks, respectively, for each player $i$ across all rounds, and $n$ is the total number of player-round observations.

Secondary metrics include:
\begin{itemize}
  \item Root mean squared error (RMSE), $L_2$. This metric gives more weight to larger errors:
    \begin{equation}
    L_2 = \sqrt{\frac{1}{n} \sum_{i=1}^{n} ( \log_2{\hat{r_i}} - \log_2{r_i} )^2}
    \end{equation}
  
  \item Kendall's Tau and Spearman's Rho~\cite{Kendall}.
    These metrics measure the correlation between the most probable order (seeding) and actual ranks.
\end{itemize}

\subsection{Properties}

\begin{figure}[h]
\centering
\includegraphics[width=0.7\textwidth]{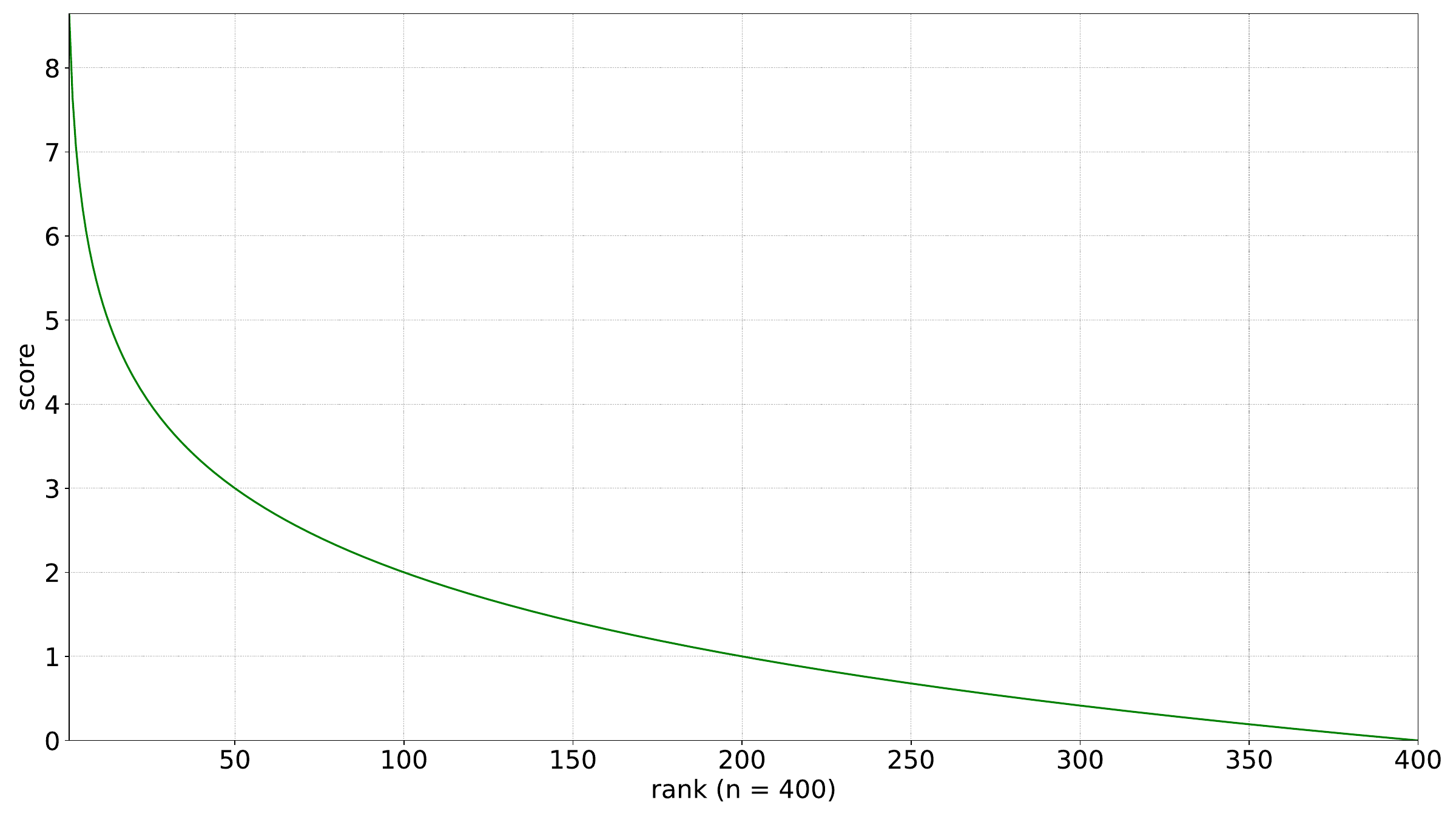}
\caption{Rank performance}
\label{fig:RankPerf}
\end{figure}

Rank performance is convex as a function of rank (Figure~\ref{fig:RankPerf}).
The total performance scores of a set of ranks is maximal when the ranks are distinct, and minimal when the ranks are tied.
Having split ties equally in actual and expected ranks, the expected ranks are at least as tied as the actual ranks.
Thus, the performances in a round have a positive or zero sum:
\begin{equation}
\sum{P} \ge 0
\end{equation}

Let $\overline{P}$ be the average performance in a round and $L_1$ the average absolute value. We find:
\begin{align}
\overline{P} &< \log_2 e - 1 \approx 0.443 \\
L_1 &< 2
\end{align}

These bounds enable meaningful cross-round comparisons and aggregation of statistics.

The regularized derivative of a player's expected performance with respect to rating is given by:
\begin{equation}
\begin{gathered}
\frac{dP}{dR} = \frac{P'}{K_0} \\
P' = \frac{1 + \sum_{j \neq i} w_j(1-w_j)}{1 + \sum_{j \neq i} w_j} \\
K_0 = \frac{400}{\log_2{10}} \approx 120
\end{gathered}
\end{equation}

For $P \approx 0$, we can solve $P = 0$ with $\Delta R = K_0 \cdot \frac{P}{P'}$.

We will use $P'$ to improve the accuracy of $\Delta R$.

\section{Proposed rating system}

We computed a performance measure for each player in a SRM.
The objective is to define rating changes $\Delta R$ that improve the prediction of future performances and allow ratings to track long-term player proficiency.

The rating system is developed incrementally, introducing one factor at a time to address a distinct aspect of performance measurement.
Each factor is theoretically motivated, and parameters are selected by minimizing the average prediction error ($L_1$) over the historical SRM dataset.

Throughout, factor selection is restricted to admissible mechanisms under which the rating scale retains a consistent interpretation as latent player proficiency.
Models that improve predictive accuracy by violating this interpretation are excluded.

\subsection{Initial factor}

With a prior ratio of $1 : 1$, a performance value $P$ outperforms the expected performance level by a factor $1 : 2^P$.
A rating difference $\Delta R$ expects a better performance by a factor $1 : 10^{\Delta R/400}$.
Equating these two expressions converts performance units to rating units:
\begin{equation}
\begin{gathered}
\Delta R = \frac{400}{\log_2{10}} P = K_0 \cdot P \\
K_0 \approx 120
\end{gathered}
\end{equation}

If we expected the same performance in the next round and had no other information, this would be an appropriate $\Delta R$.

We use this initial formulation to establish baseline statistics before customizing for SRM.

\subsection{Fixed K}

\begin{figure}[h]
\centering
\includegraphics[width=0.7\textwidth]{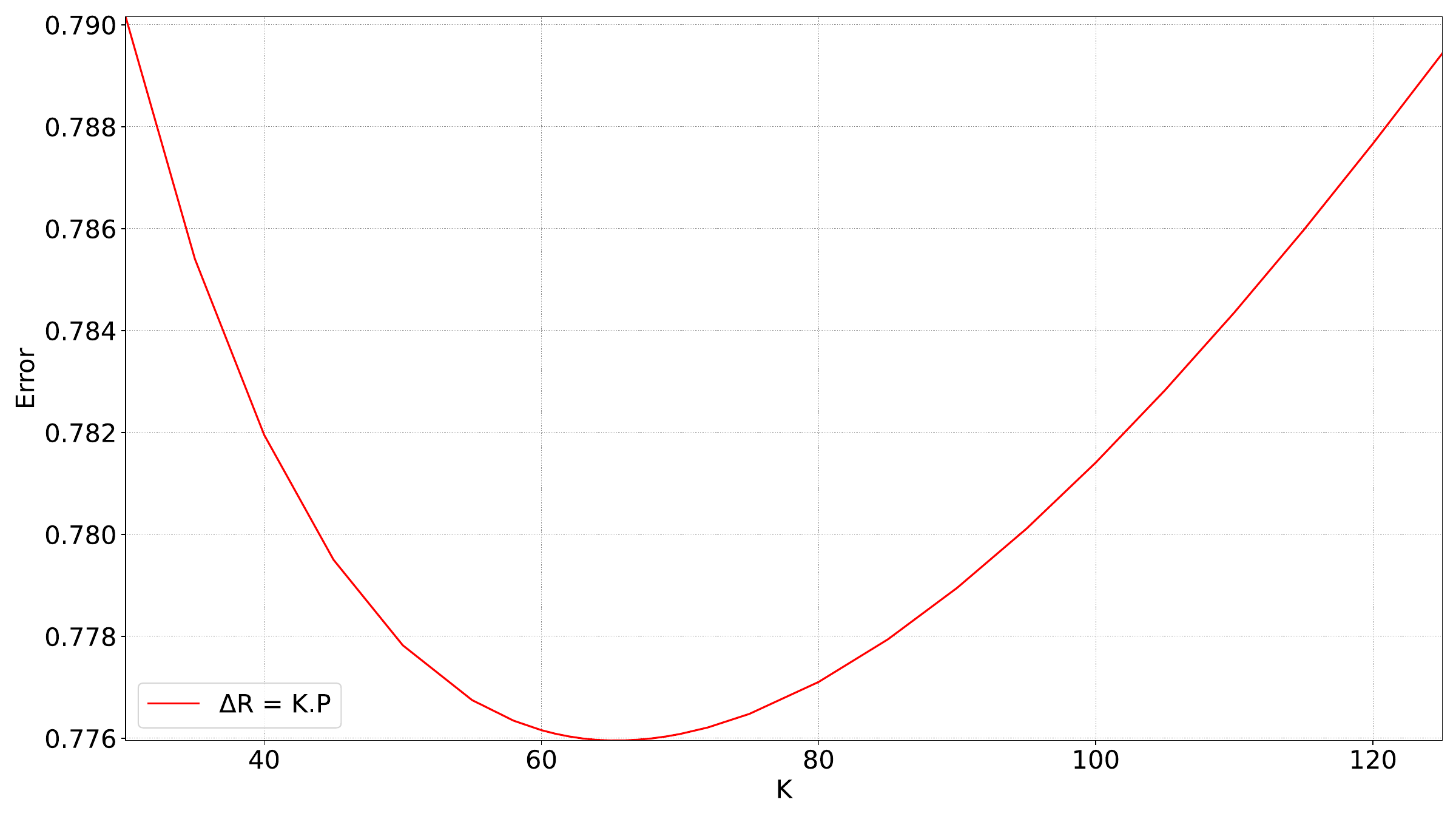}
\caption{Fixed K}
\label{fig:fixed_K}
\end{figure}

For a simple linear adjustment, we compute $\Delta R = K \cdot P$, where $K$ is chosen to minimize prediction error.
We find the most accurate choice is $K = 65$ (Figure~\ref{fig:fixed_K}).

As we add factors into the calculation of $\Delta R$, we re-optimize previous parameters to maintain accuracy, with the parameter $K$ serving as the base rate of rating change.

\subsection{Experience factor}

\begin{figure}[h]
\centering
\includegraphics[width=0.7\textwidth]{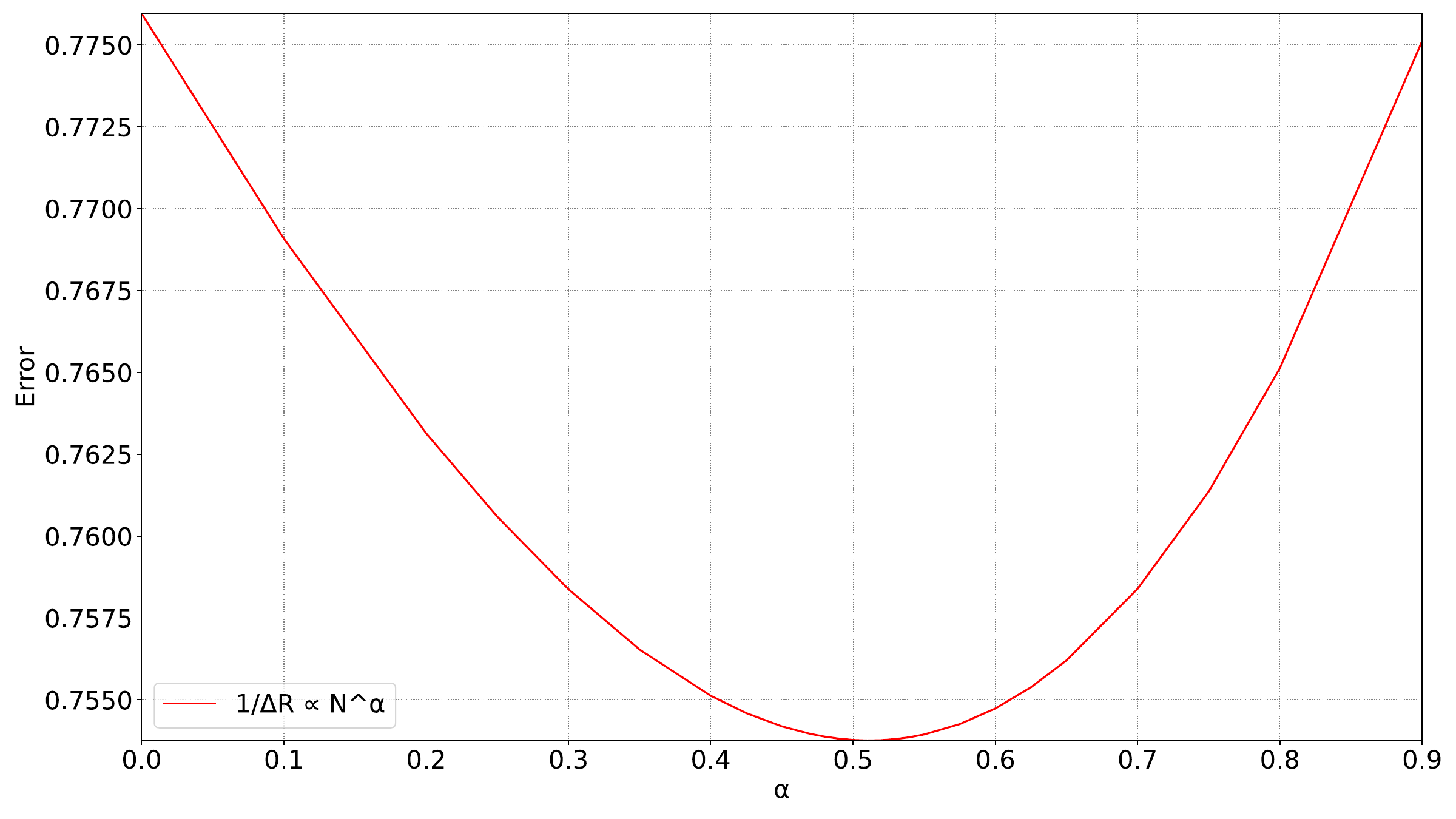}
\caption{Experience factor}
\label{fig:alpha}
\end{figure}

A player's rating represents accumulated evidence of skill, and greater experience should reduce the sensitivity of the rating to new observations.
Let $W$ be an experience-based weight, such that $\Delta R \propto 1/W$, and let $N$ denote the round number for the player.

We tested multiple functional forms, and found the best results with $W = \sqrt{N}$.
Figure~\ref{fig:alpha} shows choices of $W = N^\alpha$, with $\alpha$ a parameter.
Thus, we adopt $W = \sqrt{N}$.
\begin{equation}
\Delta R = \frac{K}{W} \cdot P
\end{equation}

This scaling moderates rating fluctuations for experienced players, improving long-term stability.

\subsection{Maximum factor}

Extreme performances predict future outcomes less reliably than performances near expectation.
We therefore limit the magnitude of $P$ to a maximum $M$ using a sigmoid function, maintaining symmetry and linearity around zero.
We find the best results with:
\[
\Delta R \propto \frac{P}{1+\frac{|P|}{M}}
\]

\begin{figure}[h]
\centering
\includegraphics[width=0.7\textwidth]{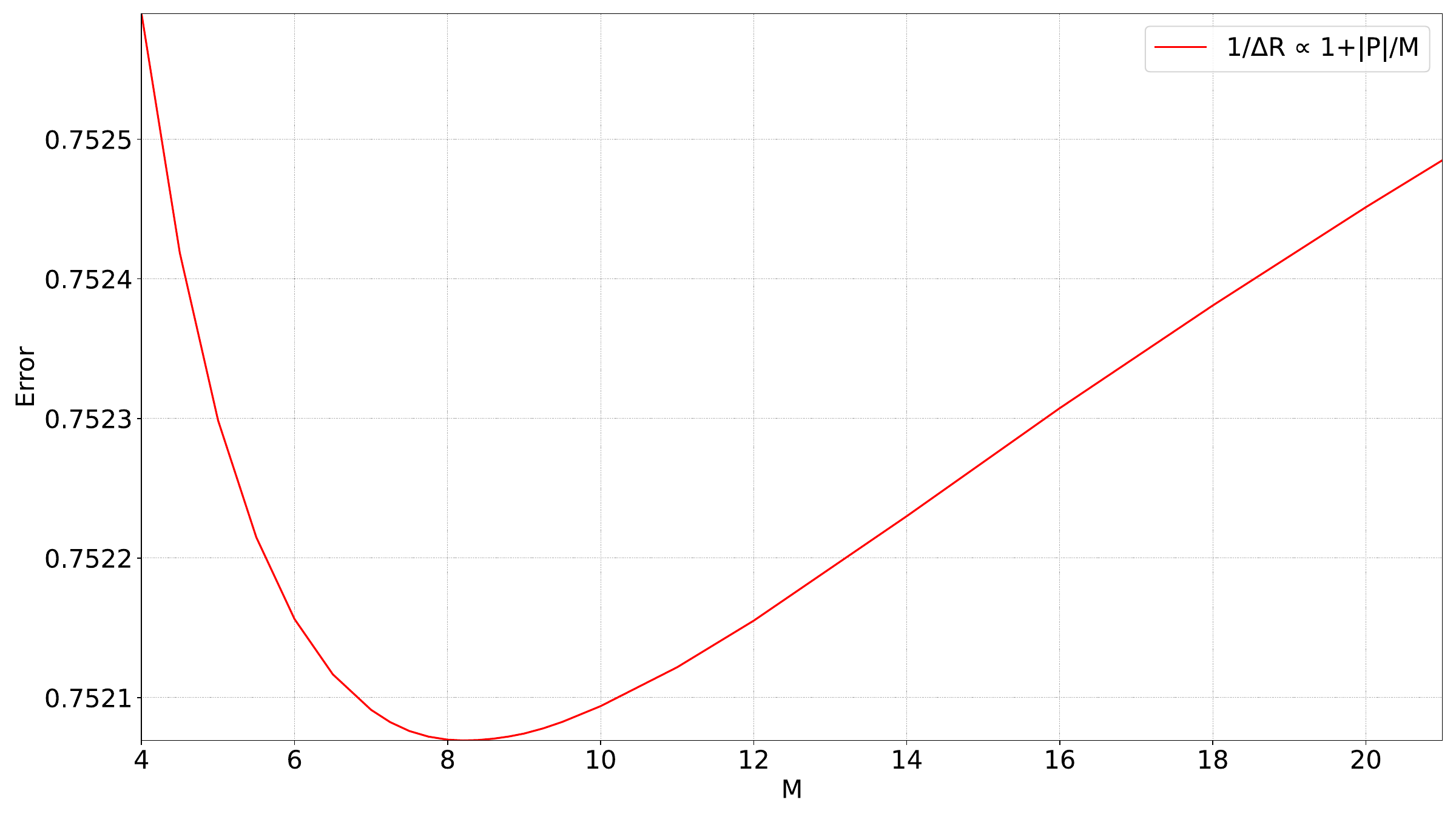}
\caption{Choice of $M$}
\label{fig:M}
\end{figure}

The most accurate $M \approx 8.2$, as shown in Figure~\ref{fig:M}.
Therefore, we set $M = 8.2$.
The rating change formula is now:
\begin{equation}
\Delta R = \frac{K}{W} \cdot \frac{P}{1+\frac{|P|}{M}}
\end{equation}

By bounding $|P|$, we reduce the impact of uncharacteristic performances that are less predictive of long-term skill.

\subsection{SRMFix}

With the inclusion of these factors, the rating change formula is similar to SRM ratings.
We will compare SRM ratings with this initial configuration, labeled \textbf{SRMFix}:

\begin{center}
SRMFix: $K = 247$; $W = \sqrt{N}$; $M = 8.2$
\end{center}

This configuration serves as a baseline for subsequent refinements.

\subsection{Competition factor}

A player's observed rank may vary for reasons unrelated to their actual skill, such as round conditions or stochastic outcomes against opponents.
We estimate this variability through $P'$, defined earlier as:
\[
P' = \frac{1 + \sum w_j(1-w_j)}{1 + \sum w_j}
\]

This expression captures the variance in expected rank and the sensitivity of performance to rank changes. The following limiting cases illustrate its interpretation:
\begin{itemize}
  \item $\lim_{\hat{r} \to 1} P' = 1$: Most impact for top players. For example, a player ranking 2nd instead of 1st ($P = -1$) could result from a single opponent having an atypical performance. Such outcomes are relatively noisy.
  
  \item $\lim_{\hat{r} \to n} P' \approx \frac{1}{n}$: Low impact for lower-ranked players. For example, a player ranking 1000th instead of 2000th ($P = 1$) would require 1000 stronger opponents to simultaneously underperform---unlikely to occur by chance. Such outcomes are more indicative of the player's own performance.
\end{itemize}

\begin{figure}[h]
\centering
\includegraphics[width=0.7\textwidth]{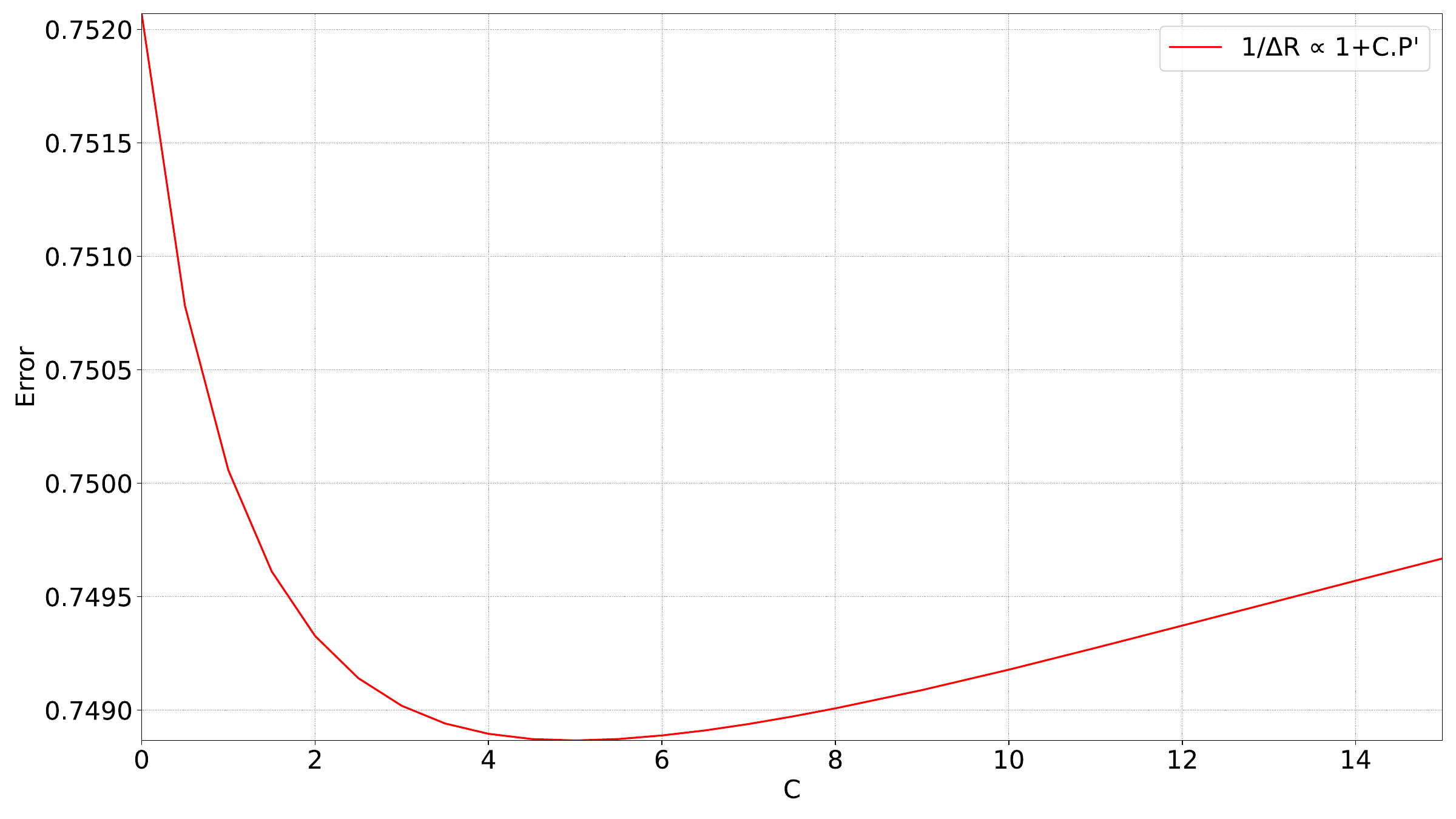}
\caption{Choice of $C$}
\label{fig:C}
\end{figure}

We compute the competition factor as:
\begin{equation}
  V = 1 + C \cdot P'
\end{equation}
where $C$ is a tunable parameter, illustrated in Figure~\ref{fig:C}.

The formula for the rating change becomes:
\begin{equation}
  \Delta R = \frac{K}{W} \cdot \frac{P}{1 + \frac{|P|}{M}} \cdot \frac{1}{V}
\end{equation}

By damping the noise present in performance measurements, this factor helps ensure that a player's underlying skill remains the primary driver of rating adjustments.

\subsection{Stability}

We have computed $\Delta R \propto P$, improving predictive accuracy post-round.
Each player is expected a performance of $0$, resulting in an expected $\Delta R = 0$.
Thus, players performing at their expected level have stable ratings.

However, several systematic effects influence rating dynamics:
\begin{itemize}
  \item Because performances have a positive sum, more rating points are won by the outperforming players than are lost by the underperforming players. The ratings have net inflation.
  \item Because players gain experience during a round, they on average have better performances after the round. Thus, some inflation better predicts future performance than deflation.
  \item When $K$ minimizes prediction error, the average rating change provides a first-order estimate of the shift in next-round expected performance of participants relative to non-participants.
\end{itemize}

This rate of inflation captures relative performance improvements during a round, but may not necessarily provide a full accounting of the players' performances yet.

Following Elo's framework, the rating system's primary purpose is to measure player proficiency~\cite[§3.5]{Elo}.
Proficiency can also change between rounds, necessitating additional factors for a more accurate rating.

The remaining factors model proficiency change and function as deflation control processes~\cite[§3.7]{Elo}.
We introduce bias-correcting adjustments that align expected and observed performance ($B$, $G$), followed by stabilizing factors that regulate rating evolution as evidence accumulates ($F$, $W_1$).

\subsection{Participation bonus}

While constrained by $\Delta R \propto P$, the expected rating change of any participant remains zero.
However, achieving the expected performance in a round is better than not participating.
Players having practiced already have better proficiencies before the round.
Thus, $\Delta R = 0$ in expectation underestimates actual proficiency gains.

We adjust the expectations to account for these improvements by simulating an increase in opponent ratings before the round.
We choose a parameter $B$ to represent the increase, then add the difference in expected performance to each player's performance:
\begin{equation}
\Delta P = \frac{B}{K_0} \cdot P'
\end{equation}

This aligns with Elo's principle~\cite[§3.53]{Elo}:
\begin{quote}
A player whose proficiency improves while he is in the pool is entitled to additional rating points, but they may not properly be taken from players whose proficiency is not changing.
\end{quote}

This factor alone provides little accuracy gains; its primary benefit emerges when combined with new-player baseline adjustment.

\subsection{New players}

A key challenge in Elo systems is the appropriate entry point for newly joining players~\cite[§3.7]{Elo}.
Previously, we assigned a constant rating $R_\text{init} = 1200$ to new players.
However, the proficiency of new players is not stationary.
As existing players improve, SRMs become increasingly difficult, raising the barrier to entry for newcomers.

Before participating, potential players have opportunities to practice on recent rounds.
Some may be experienced contestants coming from other platforms.
Thus the proficiency of new players improves over time, requiring adjustment for inflation.

After experimentation, we opt for a linear increase in $R_\text{init}$ over time:
\begin{itemize}
  \item We choose a parameter $G$, the increase in $R_\text{init}$ per year.
  \item Start with $R_\text{init} = 1200$ at the first SRM.
  \item Each round, set $R_\text{init}$ as follows:
\end{itemize}
\begin{equation}
R_\text{init}(t) = 1200 + G \cdot (t - t_0)
\end{equation}
where $t$ is the round date and $t_0$ the date of the first SRM, in years.

This adjustment allows initial ratings to reflect increasing SRM difficulty and participant proficiency over the 23-year SRM history.

\subsection{Frequency factor}

A player's proficiency can change over time since the last measurement.
Thus, players more recently rated tend to have more accurate ratings.
Frequent participants have more stable proficiencies between rounds than infrequent participants.
Therefore, we may further stabilize the ratings of recent and frequent participants for accuracy:
\begin{itemize}
  \item Define $N_r$ as a player's effective number of recent rounds, where each round contributes 1 initially and decays exponentially over time.
  \item Let $D$ be the half-life of a round’s relevance, in days.
  \item The decay rate is $\lambda = e^{-\ln(2) / D}$.
  \item In a round played $\Delta t$ days after the previous one, update:
    \begin{equation}
    N_r = 1 + N_{r,\text{prev}} \cdot \lambda^{\Delta t}.
    \end{equation}
  \item The frequency factor is $F = N_r^{\gamma}$, with parameter $\gamma$.
\end{itemize}

The rating change is stabilized by:
\begin{equation}
\Delta R_{\text{adjusted}} = \Delta R \cdot \frac{1}{F}
\end{equation}
reducing volatility for consistently active players while allowing larger updates after periods of inactivity.

\subsection{Revision of experience factor}

Earlier we defined $W = \sqrt{N}$ to stabilize ratings based on cumulative experience.
However, $N$ and $N_r$ are highly correlated for active players, limiting accuracy gains from $F$.
We revise the formula to reduce this overlap:

\begin{equation}
W = \sqrt{1 + W_1 (N-1)}
\end{equation}
where $W_1$ reduces the weight of previous rounds relative to the current round.

This allows $W$ and $F$ to capture complementary aspects:
\begin{itemize}
  \item $W$: Reflects total accumulated evidence, grows slowly with career length
  \item $F$: Reflects recent measurement reliability, decays rapidly with inactivity
\end{itemize}

\subsection{EloSRM}

Optimizing all parameters jointly, we obtain approximately:
$K = 648$, $C = 3.9$, $M = 4.4$, $B = 42$, $G = 51$, $W_1 = .20$, $\gamma = 0.46$, $D = 112$.

Exact values and implementation are given in the source code.

With these components, we define \textbf{EloSRM}, a rating system for SRMs intended to measure player proficiency over time.

Table~\ref{tab:elosrm_components} summarizes the components of EloSRM in the order they are applied.
Together, they fully specify the rating update algorithm.

\begin{table}[H]
\centering
\renewcommand{\arraystretch}{1.3}
\makebox[\textwidth]{%
\begin{tabular}{@{} l >{\raggedright\arraybackslash}p{4.2cm} p{6.3cm} @{}}
  \toprule
  \textbf{Component} & \textbf{Definition} & \textbf{Role} \\
  \midrule
  Performance & \mbox{$P = \log_2{\hat{r}} - \log_2{r}$} & Converts ranks to wins-equivalent score \\
  \addlinespace
  Maximum factor & \mbox{$P_M = \dfrac{P}{1 + |P|/M}$} & Bounds influence of extreme performances \\
  \addlinespace
  Performance sensitivity & \mbox{$P' = \dfrac{1 + \sum w_j(1-w_j)}{1 + \sum w_j}$} & Quantifies rating gradient and noise sensitivity \\
  \addlinespace
  Participation bonus & \mbox{$P_B = \dfrac{B}{K_0}\,P'$} & Accounts for proficiency gains associated with participation and practice \\
  \addlinespace
  Adjusted performance & \mbox{$P_A = P_M + P_B$} & Effective performance used for rating updates \\
  \addlinespace
  Experience factor & \mbox{$W = \sqrt{1 + W_1 (N-1)}$} & Stabilizes ratings as evidence accumulates \\
  \addlinespace
  Competition factor & \mbox{$V = 1 + C\,P'$} & Damps measurement noise \\
  \addlinespace
  Frequency factor & \mbox{$F = N_r^{\gamma}$} & Accounts for participation frequency \\
  \addlinespace
  Rating change & \mbox{$\Delta R = \dfrac{K}{W\,V\,F}\,P_A$} & Per-round rating update \\
  \addlinespace
  New-player baseline & \mbox{$R_\text{init}(t) = 1200 + G\,(t - t_0)$} & Adjusts entry rating over time \\
  \bottomrule
\end{tabular}
}
\caption{EloSRM components}
\label{tab:elosrm_components}
\end{table}

\section{Experimental results}

Results include all rated SRM up to May 2024.

We first compare our SRMFix implementation to the original SRM ratings.
Table~\ref{tab:srmfix_srm_players} shows the average $\Delta R$, performance, and prediction error, using our definitions.
\begin{itemize}
  \item The first row is our primary metric, averaging all participants.
  \item Subsequent rows categorize players by experience levels.
  \item The ``Existing'' row includes all players beyond their first round.
  \item Existing players in each division.
  \item Within each division, the top and bottom half ranks.
\end{itemize}

\begin{table}[H]
\centering
\makebox[\textwidth]{%
\begin{tabular}{l r r r r r r r}
  \toprule
   & & \multicolumn{2}{c}{\textbf{ΔR}} & \multicolumn{2}{c}{\textbf{perf}} & \multicolumn{2}{c}{\textbf{err}} \\
  \cmidrule(lr){3-4} \cmidrule(lr){5-6} \cmidrule(lr){7-8}
  \textbf{Players} & \textbf{ratings} & \textbf{SRMFix} & \textbf{SRM} & \textbf{SRMFix} & \textbf{SRM} & \textbf{SRMFix} & \textbf{SRM} \\
  \midrule
  All               & 817969 &  17.6 & -20.9 & 0.221 & 0.297 & \textbf{0.7521} & 0.8273 \\
  First round       &  82174 &  54.5 &-184.8 & 0.359 &-0.362 & \textbf{0.8038} & 1.0691 \\
  2--7 rounds       & 209177 &  29.8 & -24.8 & 0.300 & 0.247 & \textbf{0.6698} & 0.7106 \\
  8--24 rounds      & 223176 &  12.2 &  10.9 & 0.241 & 0.509 & \textbf{0.7470} & 0.8177 \\
  25--74 rounds     & 199983 &   4.7 &   4.5 & 0.161 & 0.393 & \textbf{0.7662} & 0.8232 \\
  75--199 rounds    &  88655 &   0.5 &  -0.7 & 0.040 & 0.283 & \textbf{0.8624} & 0.9009 \\
  200+ rounds       &  14804 &  -0.7 &  -1.8 &-0.044 & 0.243 & \textbf{0.8529} & 0.8929 \\
  Existing          & 735795 &  13.5 &  -2.6 & 0.206 & 0.371 & \textbf{0.7463} & 0.8003 \\
  Division 1        & 405158 &   9.3 &  -1.8 & 0.162 & 0.241 & \textbf{0.6928} & 0.7200 \\
  Division 2        & 330637 &  18.6 &  -3.6 & 0.259 & 0.529 & \textbf{0.8118} & 0.8987 \\
  D1 H1             & 202579 &  30.0 &  51.8 & 0.621 & 0.791 & \textbf{0.9725} & 1.0375 \\
  D1 H2             & 202579 & -11.3 & -55.5 &-0.296 &-0.309 & 0.4132 & \textbf{0.4025} \\
  D2 H1             & 165319 &  64.0 &  55.2 & 0.890 & 1.348 & \textbf{1.1448} & 1.3832 \\
  D2 H2             & 165318 & -26.8 & -62.5 &-0.371 &-0.289 & 0.4789 & \textbf{0.4142} \\
  \bottomrule
\end{tabular}
}
\caption{Player statistics, SRMFix vs SRM}
\label{tab:srmfix_srm_players}
\end{table}

Table~\ref{tab:srmfix_srm_rounds} shows round statistics.

For each metric, we compute the fraction of rounds where SRMFix better predicted the result than SRM ratings, splitting ties equally.
Table~\ref{tab:srmfix_srm_rounds} shows the percentages.
Rows are grouped by division, then by round size.

\begin{table}[H]
\centering
\begin{tabular}{l r r r r r}
  \toprule
    & & \multicolumn{4}{c}{\textbf{SRMFix > SRM (\%)}} \\
  \cmidrule(lr){3-6}
  \textbf{Rounds} & \textbf{\#} & 
  \textbf{\makebox[1.2cm][r]{\bm{$L_1$}}} & 
  \textbf{\makebox[1.2cm][r]{\bm{$L_2$}}} & 
  \textbf{\makebox[1.2cm][r]{\textbf{Tau}}} & 
  \textbf{\makebox[1.2cm][r]{\textbf{Rho}}} \\
  \midrule
  All                 & 2182 & 86.5 & 87.0 & 83.0 & 82.4 \\
  Division 1          & 1356 & 79.5 & 81.6 & 73.4 & 72.6 \\
  Division 2          &  826 & 97.9 & 95.9 & 98.8 & 98.5 \\
  2--16 players       &  167 & 54.8 & 53.0 & 51.5 & 52.4 \\
  17--99 players      &  323 & 68.4 & 67.8 & 64.6 & 62.4 \\
  100--199 players    &  394 & 88.5 & 85.4 & 80.1 & 79.1 \\
  200--399 players    &  469 & 92.8 & 94.0 & 88.7 & 88.3 \\
  400--599 players    &  317 & 93.1 & 96.5 & 91.5 & 91.8 \\
  600--799 players    &  268 & 96.3 & 98.9 & 95.9 & 95.9 \\
  800+ players        &  243 & 97.9 & 99.6 & 97.9 & 97.1 \\
  \bottomrule
\end{tabular}
\caption{Round statistics, SRMFix vs SRM}
\label{tab:srmfix_srm_rounds}
\end{table}

\FloatBarrier

Next, we evaluate our EloSRM implementation in contrast to SRMFix.
Tables~\ref{tab:elosrm_srmfix_players} and~\ref{tab:elosrm_srmfix_rounds} show the player and round statistics, respectively.

Table~\ref{tab:ratings} summarizes the statistics and incremental contributions of each factor.

\begin{table}[H]
\centering
\makebox[\textwidth]{%
\begin{tabular}{l r r r r r r r}
  \toprule
    & & \multicolumn{2}{c}{\textbf{ΔR}} & \multicolumn{2}{c}{\textbf{perf}} & \multicolumn{2}{c}{\textbf{err}} \\
  \cmidrule(lr){3-4} \cmidrule(lr){5-6} \cmidrule(lr){7-8}
  \textbf{Players} & \textbf{ratings} & \textbf{EloSRM} & \textbf{SRMFix} & \textbf{EloSRM} & \textbf{SRMFix} & \textbf{EloSRM} & \textbf{SRMFix} \\
  \midrule
    All               & 817969 & 29.3 & 17.6 & 0.218 & 0.221 & \textbf{0.7449} & 0.7521 \\
    First round       &  82174 &107.8 & 54.5 & 0.458 & 0.359 & \textbf{0.7980} & 0.8038 \\
    2--7 rounds       & 209177 & 45.4 & 29.8 & 0.265 & 0.300 & \textbf{0.6626} & 0.6698 \\
    8--24 rounds      & 223176 & 16.4 & 12.2 & 0.194 & 0.241 & \textbf{0.7376} & 0.7470 \\
    25--74 rounds     & 199983 &  8.1 &  4.7 & 0.174 & 0.161 & \textbf{0.7607} & 0.7662 \\
    75--199 rounds    &  88655 &  3.6 &  0.5 & 0.083 & 0.040 & \textbf{0.8555} & 0.8624 \\
    200+ rounds       &  14804 &  1.6 & -0.7 &-0.011 &-0.044 & \textbf{0.8459} & 0.8529 \\
    Existing          & 735795 & 20.5 & 13.5 & 0.191 & 0.206 & \textbf{0.7390} & 0.7463 \\
    Division 1        & 405158 & 15.7 &  9.3 & 0.167 & 0.162 & \textbf{0.6869} & 0.6928 \\
    Division 2        & 330637 & 26.5 & 18.6 & 0.222 & 0.259 & \textbf{0.8028} & 0.8118 \\
    D1 H1             & 202579 & 35.4 & 30.0 & 0.628 & 0.621 & \textbf{0.9636} & 0.9725 \\
    D1 H2             & 202579 & -4.0 &-11.3 &-0.295 &-0.296 & \textbf{0.4101} & 0.4132 \\
    D2 H1             & 165319 & 64.4 & 64.0 & 0.840 & 0.890 & \textbf{1.1050} & 1.1448 \\
    D2 H2             & 165318 &-11.4 &-26.8 &-0.397 &-0.371 & 0.5006 & \textbf{0.4789} \\
  \bottomrule
\end{tabular}
}
\caption{Player statistics, EloSRM vs SRMFix}
\label{tab:elosrm_srmfix_players}
\end{table}

\begin{table}[H]
\centering
\begin{tabular}{l r r r r r}
  \toprule
    & & \multicolumn{4}{c}{\textbf{EloSRM > SRMFix (\%)}} \\
  \cmidrule(lr){3-6}
  \textbf{Rounds} & \textbf{\#} & 
  \textbf{\makebox[1.2cm][r]{\bm{$L_1$}}} & 
  \textbf{\makebox[1.2cm][r]{\bm{$L_2$}}} & 
  \textbf{\makebox[1.2cm][r]{\textbf{Tau}}} & 
  \textbf{\makebox[1.2cm][r]{\textbf{Rho}}} \\
  \midrule
    All                 & 2182 & 76.1 & 73.7 & 72.6 & 68.2 \\
    Division 1          & 1356 & 72.2 & 72.5 & 71.4 & 69.8 \\
    Division 2          &  826 & 82.7 & 75.5 & 74.5 & 65.7 \\
    2--16 players       &  167 & 52.4 & 54.8 & 53.9 & 55.4 \\
    17--99 players      &  323 & 63.2 & 63.5 & 59.4 & 59.4 \\
    100--199 players    &  394 & 76.3 & 77.5 & 74.9 & 72.2 \\
    200--399 players    &  469 & 81.4 & 74.8 & 75.7 & 70.4 \\
    400--599 players    &  317 & 77.6 & 77.9 & 77.6 & 72.2 \\
    600--799 players    &  268 & 82.1 & 77.6 & 78.4 & 71.6 \\
    800+ players        &  243 & 90.9 & 81.9 & 80.2 & 69.1 \\
  \bottomrule
\end{tabular}
\caption{Round statistics, EloSRM vs SRMFix}
\label{tab:elosrm_srmfix_rounds}
\end{table}

\begin{table}[H]
\centering
\makebox[\textwidth]{%
\begin{tabular}{l r r r r @{\hspace{1.5em}} r r r @{\hspace{1.5em}} c @{\hspace{0.3em}} c @{\hspace{0.3em}} c}
  \toprule
    & \multicolumn{4}{c}{\textbf{error}} & \multicolumn{3}{c}{\bm{$\Delta R$}} & \multicolumn{3}{c}{\textbf{R}} \\
  \cmidrule(r{1.5em}){2-5} \cmidrule(r{1.5em}){6-8} \cmidrule(){9-11}
  \textbf{Rating} & \bm{$L_1$} & \bm{$L_2$} & \textbf{Tau} & \textbf{Rho} & \bm{$\mu$} & \bm{$\sigma$} & \textbf{max} & \textbf{init} & \textbf{median} & \textbf{max} \\
  \midrule
    EloSRM   & 0.7449 & 1.1282 & 0.419 & 0.571 & 29.3 & 105 & 1385 & 2370 & 2605 & 4776 \\
    SRMFix   & 0.7521 & 1.1355 & 0.413 & 0.565 & 17.6 &  98 & 1096 & 1200 & 1520 & 3810 \\
    SRM      & 0.8273 & 1.2219 & 0.294 & 0.398 &-20.9 & 131 & 1185 & 1200 & 1198 & 4107 \\
    initial  & 0.7878 & 1.1879 & 0.395 & 0.544 & 22.1 & 141 & 1202 & 1200 & 1456 & 4663 \\
    K        & 0.7760 & 1.1702 & 0.394 & 0.543 & 14.6 &  75 &  642 & 1200 & 1369 & 3962 \\
    W        & 0.7538 & 1.1370 & 0.412 & 0.564 & 18.8 & 103 & 1781 & 1200 & 1515 & 3816 \\
    M        & 0.7521 & 1.1355 & 0.413 & 0.565 & 17.6 &  98 & 1096 & 1200 & 1520 & 3810 \\
    C        & 0.7489 & 1.1322 & 0.415 & 0.568 & 19.7 & 103 & 1617 & 1200 & 1550 & 3785 \\
    B        & 0.7487 & 1.1328 & 0.416 & 0.568 & 20.9 & 100 & 1518 & 1200 & 1564 & 3853 \\
    G        & 0.7473 & 1.1306 & 0.417 & 0.569 & 24.5 &  99 & 1491 & 2042 & 2257 & 4445 \\
    F        & 0.7458 & 1.1289 & 0.418 & 0.570 & 27.3 & 107 & 1451 & 2156 & 2411 & 4606 \\
    $W_1$    & 0.7449 & 1.1282 & 0.419 & 0.571 & 29.3 & 105 & 1385 & 2370 & 2605 & 4776 \\
  \bottomrule
\end{tabular}
}
\caption{Statistics, each factor}
\label{tab:ratings}
\end{table}

\FloatBarrier

\section{Interpretation}

SRMFix improves prediction accuracy over traditional SRM ratings. Rank correlations (Kendall's $\tau$, Spearman's $\rho$) show systematic gains despite not being directly optimized, indicating genuine improvements in ordering predictions.

EloSRM extends the prediction model by incorporating proficiency measurement.
Rating progressions show large initial gains for newcomers tapering to small increases for veterans, consistent with skill development patterns Elo observed in chess~\cite[§3.8]{Elo}.
However, experienced players (200+ rounds) show continued positive changes beyond typical chess plateaus, which may reflect evolving SRM difficulty, self-selection of long-term participants, or other factors.

While the models could be further refined, these patterns suggest EloSRM provides a stable measure of SRM proficiency over time.

Our metrics measure relative within-round performance and cannot identify absolute proficiency calibration from rankings alone.
The inflation rate represents one plausible solution among those compatible with the data.

\section{Adherence to Elo principles}

EloSRM incorporates the structural requirements of an Elo-based rating system:
\begin{itemize}
  \item \textbf{Rating-difference-based expectations:} Expected performance uses the logistic form $w_{ij} = 1/(1 + 10^{(R_i - R_j)/400})$, the probability model adopted in modern Elo implementations~\cite[§1.15, §8.73]{Elo}\cite{FIDE}.
  
  \item \textbf{Continuous measurement:} Ratings update after each round via $\Delta R = f(P)$, where performance $P$ is measured against expectations, following the structure of Elo's continuous rating formula $R_n = R_o + K(W - W_e)$~\cite[§1.61]{Elo}.
  
  \item \textbf{Deflation control:} The participation bonus ($B$), new-player baseline ($G$), and frequency factor ($F$) function as deflation-control mechanisms~\cite[§3.7]{Elo}.
\end{itemize}

As in classical Elo systems, rating integrity is evaluated empirically through monitoring of rating behavior over time, and may require periodic adjustment as the player pool evolves~\cite[§3.5, §3.6, §3.8]{Elo}.

\section{Conclusion}

We have presented an Elo-based rating system adapted to single-round, multi-player programming contests.
Ranked outcomes are incorporated through a score-equivalent performance measure, with additional adjustments chosen to maintain predictive accuracy and rating-scale integrity.
Although the specific forms and parameters are tailored to Topcoder Single Round Matches, the underlying approach is applicable to similar ranked contests.

\section{Availability}

The EloSRM implementation, dataset, and code reproducing the main results are available under the MIT license in the project repository~\cite{EloSRM_repo}. Additional experimental results and visualizations are posted on the project website~\cite{EloSRM_web}.

\section{Acknowledgements}

We thank the competitive programming community for the historical contest data and the participants whose performances made this analysis possible.
We thank Dmitry Kamenetsky for contributions to initial document preparation and editing.

\bibliographystyle{unsrt}
\bibliography{elosrm}

\begin{appendices}

\section{Source code}
\begin{lstlisting}

//
// EloSRM rating system for Topcoder SRM
// https://github.com/batty999/EloSRM
// (c) 2019-2026 Fred Batty
//

#include <cmath>
#include <vector>


namespace EloSRM
{
    // Constants
    const int VERSION = 7;
    const double EloScale = M_LN10 / 400;
    const double K0 = 400 * M_LN2 / M_LN10;
    const double R0 = 1200;                     // Initial rating

    // System parameters
    const double K = 648.3147599935407;         // Base K-factor
    const double C = 3.8884120557511483;        // Competition factor
    const double M = 4.44015774770823;          // Max performance
    const double B = 41.84027892146161;         // Performance bonus
    const double W1 = 0.20058285315948782;      // Experience weight
    const double D = 112.17636272246507;        // Recent period
    const double gamma = 0.4599945496035186;    // Frequency factor
    const double G = 50.67198262989024;         // Inflation per year

    // Runtime computed values
    time_t t0;                                  // First round date
    double R_init = R0;                         // adjusted for inflation
    double B2;
    double lambda;
    double G_sec;

    struct Player {
        int num_ratings = 0;
        double rating = R_init;
        double recent_rounds = 0;
        time_t last_round = 0;
    };

    struct Result {
        Player* player;
        double score;
        double rate;            // 10 ** rating / 400
        double delta_r;
        double perf;
    };

    void rateDivision(Result* results, int n) {
#pragma omp parallel for
        for (int i = 0; i < n; i++) {
            Player* pi = results[i].player;
            double si = results[i].score;
            double ri = results[i].rate;
            double erank = 1, arank = 1;
            double mu = 1, var = 1;
            for (int j = 0; j < n; j++) {
                if (j == i) continue;
                double sj = results[j].score;
                double rj = results[j].rate;

                double wj = rj / (ri + rj);     // win probability
                mu += wj;
                var += wj * (1 - wj);
                if (si == sj) {
                    erank += .5;
                    arank += .5;
                }
                else {
                    erank += wj;
                    arank += si < sj;
                }
            }
            double perf = log(erank / arank) * M_LOG2E;
            double perf1 = var / mu;

            double pa = perf * M / (M + abs(perf));
            pa += B2 * perf1;
            double ef = sqrt(1. + pi->num_ratings * W1);
            double cf = 1 + C * perf1;
            double ff = pow(pi->recent_rounds, gamma);
            double w = ef * cf * ff;
            double delta_r = K * pa / w;

            results[i].delta_r = delta_r;
            results[i].perf = perf;
        }
    }

\end{lstlisting}
\pagebreak
\begin{lstlisting}

    void rateRound(std::vector<Result>& results, time_t round_time) {
        R_init = R0 + G_sec * (round_time - t0);
        int n = (int)results.size();
        for (int i = 0; i < n; i++) {
            Result* ri = &results[i];
            Player* pi = ri->player;
            if (!pi->num_ratings) {
                pi->rating = R_init;
                pi->recent_rounds = 1;
            }
            else {
                auto t_diff = round_time - pi->last_round;
                double decay = exp(lambda * t_diff);
                pi->recent_rounds = 1 + pi->recent_rounds * decay;
            }
            ri->rate = exp(pi->rating * EloScale);
        }
        rateDivision(&results[0], n);

        for (int i = 0; i < n; i++) {
            Player* pi = results[i].player;
            pi->num_ratings++;
            pi->rating += results[i].delta_r;
            pi->last_round = round_time;
        }
    }

    void init(time_t first_round_time) {
        t0 = first_round_time;
        R_init = R0;
        G_sec = G / (365.25 * 24 * 3600);
        B2 = B / K0;
        lambda = -M_LN2 / (D * 24 * 3600);
    }
}

\end{lstlisting}

\pagebreak
\section{Rating progressions}

\begin{figure}[h!]
\centering
\includegraphics[width=0.75\textwidth]{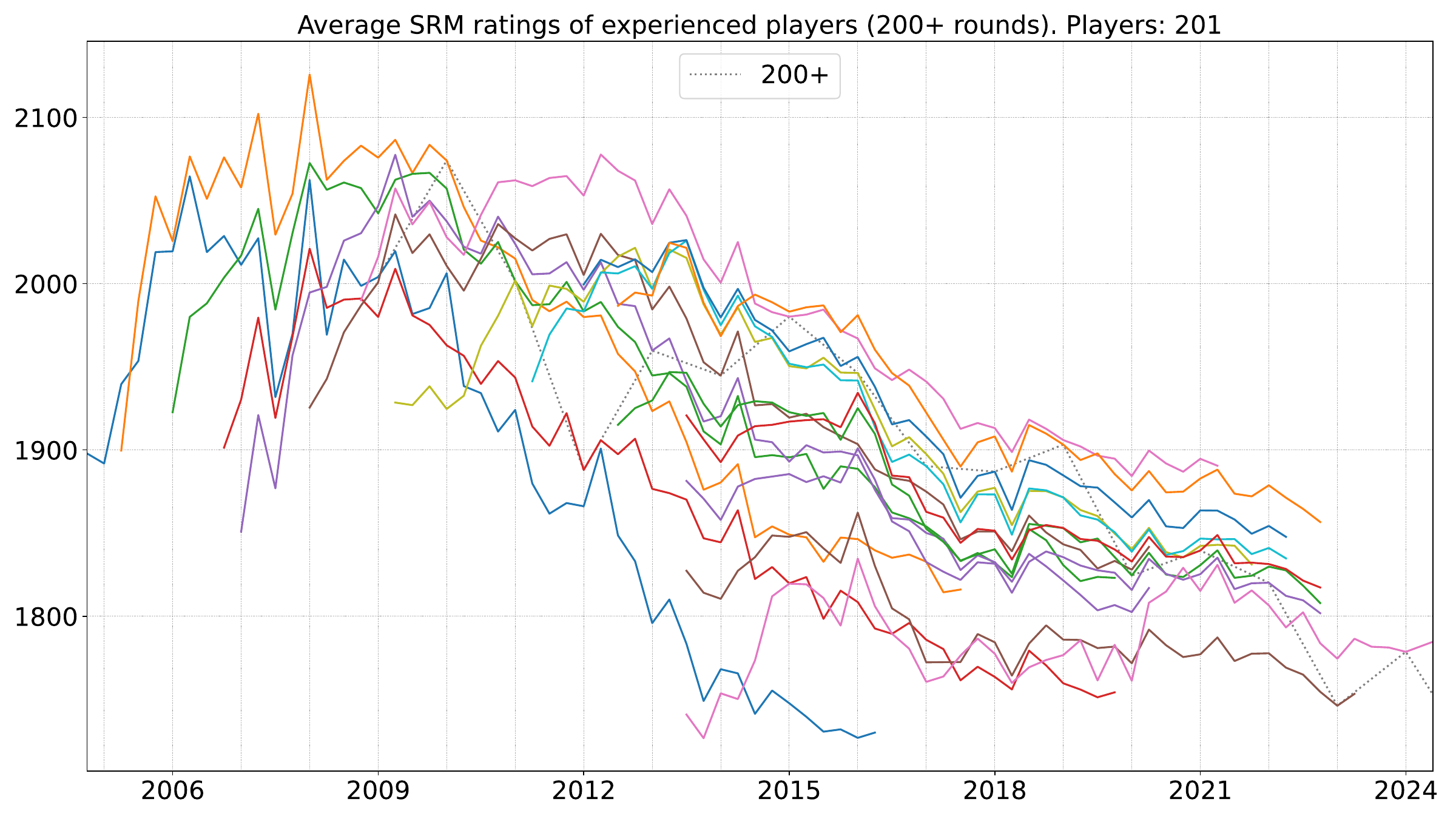}
\end{figure}

\begin{figure}[h!]
\centering
\includegraphics[width=0.75\textwidth]{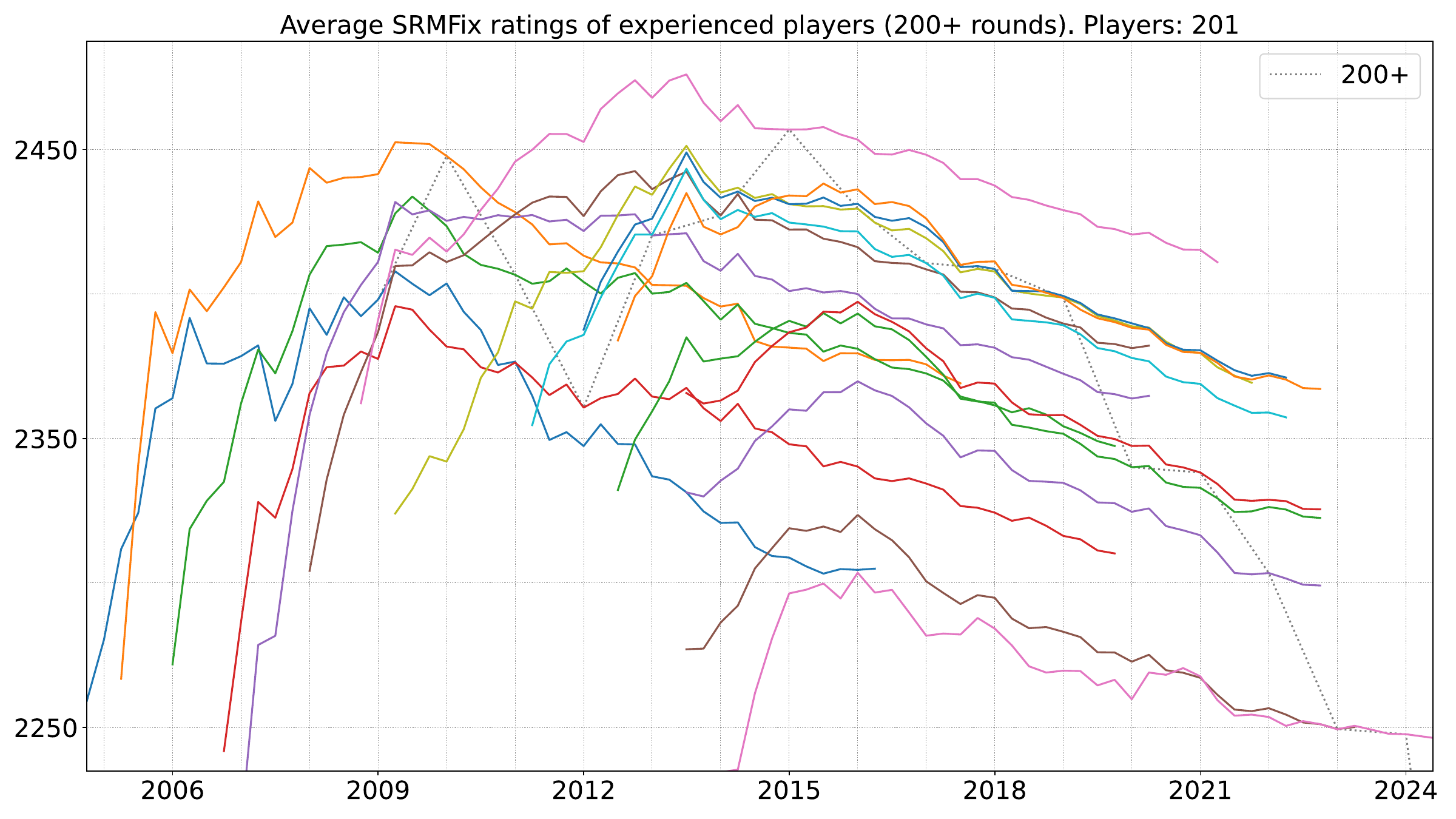}
\end{figure}

\begin{figure}[h!]
\centering
\includegraphics[width=0.75\textwidth]{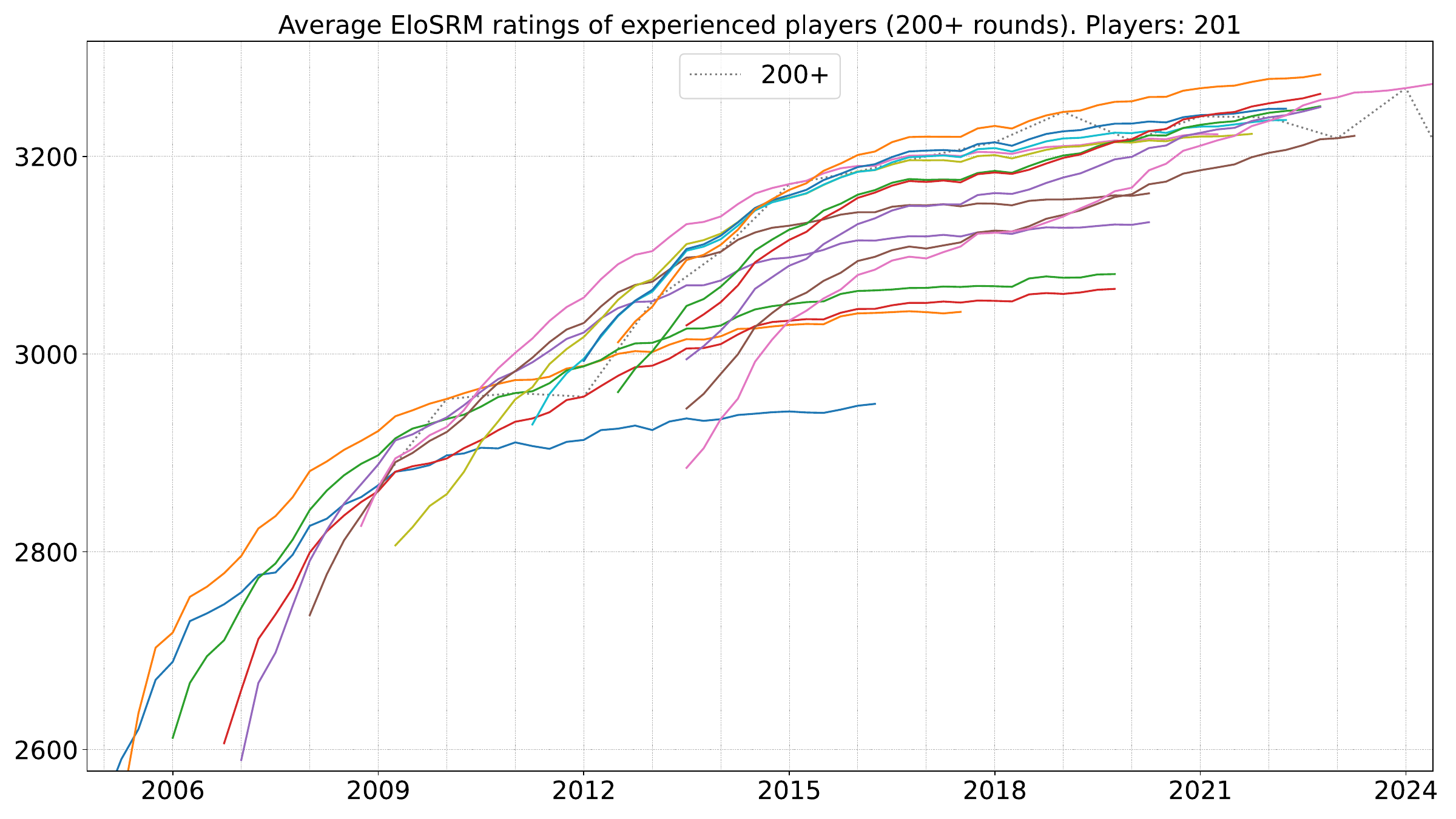}
\end{figure}
\end{appendices}

\end{document}